# Sensitiveness of the ratio between monovacancy and bulk positron lifetimes to the approximations used in the calculations: periodic behavior

J. M. Campillo-Robles[1], E. Ogando[2] and F. Plazaola[3]

1 Oinarrizko Zientziak Saila, Mondragon Goi Eskola Politeknikoa, Mondragon Unibertsitatea, 20500 Arrasate, Basque Country.

2 IZO-SGI. Servicios Generales de Investigación, Rectorado, UPV/EHU, 48080 Bilbao.

3 Elektrizitatea eta Elektronika Saila, Zientzia eta Teknologia Fakultatea, UPV/EHU, 48080 Bilbao.

E-mail: jmcampillo@mondragon.edu, edu.ogando@ehu.es , fernando@we.lc.ehu.es .

**Abstract.** Positron lifetimes have been calculated in bulk and monovacancies for most of the elements of the periodic table. Self-consistent and non-self-consistent schemes have been used for the calculation of the electronic structure in the solid, as well as different parameterizations for the positron enhancement factor and correlation energy. The ratio between the monovacancy and bulk lifetimes has been analyzed. This ratio shows a periodic behaviour with atomic number in all the calculation methods and it is in agreement with selected experimental data. The ratio shows, in contradiction to previous assumptions, sensitiveness to the approximations used in the calculations. This extensive work has allowed us to study and enlighten features of the theory and computing methods broadly used nowadays in simulating, studying and understanding positronic parameters.

## 1. Introduction

In the last years, systematic calculations of positron characteristics of the elements of the periodic table have been performed using Density Functional Theory (DFT) [1, 2]. The effort made to calculate systematically the annihilation parameters is important to deepen into the knowledge of the calculation methods, improving the theoretical background required for a good interpretation of the experimental data. In this way, recently interesting efforts have been made dealing with momentum distributions of annihilating electron-positron pairs and the energetic of positron states trapped at vacancies [3, 4]. They showed that in the case of ratios of Doppler spectra between different systems the use of a state-dependent enhancement scheme leads to better results than a position-dependent enhancement factor.

Nevertheless, changes in calculated positron parameters, when they are given as ratios between values for defects and those for the perfect lattice, have been taken to be quite insensitive to the details of the theory. For example, the positron lifetime ratio between monovacancy and bulk lifetimes, $\tau_v/\tau_b$, which reflects the degree of the localization of the positron at vacancies, has been used for comparing different calculation methods with each other and with the experimental values [5, 6]. Indeed, Barbiellini, Korhonen and coworkers calculated positron lifetimes in bulk systems with different enhancement factors, and proposed a scaling factor to relate them with the experimental values. As a result, they conclude that the $\tau_v/\tau_b$ ratio is not very sensitive to the enhancement factor used in lifetime calculation [5, 6]. In the case of the Doppler lineshape measurements, which are very valuable in the case of thin films, the lineshape is usually analyzed using the so-called $S$ and $W$ lineshape parameters. Analyses of the changes in the $W$ parameter relative to its bulk value turned out to be very useful in the identification of defects [7, 8]. But to support defect identification, calculated positron annihilation rates with different atom electrons (core, valence and d electrons) were compared with the measured ones, using the ratio between calculations in vacancies and bulk [6, 9, 10].

In this work, we analyze the behavior of the $\tau_v/\tau_b$ ratio in most of the elements of the periodic table using different enhancement factors in order to ascertain the validity of the previously cited assumptions about the insensitiveness of positron parameters defect/bulk ratio to the details of the theory.

## 2. Calculation Method

Positron lifetimes for bulk and monovacancies of most of the elements of the periodic table have been calculated. Our computational methods are described in more detail in ref. 1, here we will only explain the main features. We have used two schemes for treating the electronic densities in calculations. First, we have used a non-self-consistent scheme, the atomic superposition approximation of Puska and Nieminen (AT-SUP) [11]. Moreover, we have used a self-consistent scheme, the linear muffin-tin orbital method within the atomic-spheres approximation (LMTO-ASA) [12, 13]. The exchange and correlation effects of the positron-electrons system have been simulated using three different approximations:

(a) Within the local density approximation, the parameterization derived by Boroński and Nieminen [14]. From now it will be named as BN.

(b) Within the local density approximation, where the correlation energy of Boroński and Nieminen [14] is used, and also an enhancement factor [15] based on the values of Arponen and Pajanne [16]. This approximation will be labelled as LDA.

(c) Within the general gradient approximation, the equations obtained by Barbiellini et al. [15]. This scheme will be named as GGA.

No atomic relaxation has been taken into account in the atomic positions.

## 3. Results and discussion

For most of the elements of the periodic table bulk and monovacancy positron lifetimes have been calculated using AT-SUP and LMTO-ASA methods. Calculated lifetimes can be found in ref. 1. Using these values, we have obtained the ratio between monovacancy and bulk lifetimes, $\tau_v/\tau_b$, of each element. In Table 1 we have expressed the $\tau_v/\tau_b$ ratios calculated within AT-SUP using BN, LDA and GGA approximations. In Table 2, we have collected the $\tau_v/\tau_b$ ratios of LMTO-ASA using BN and GGA approximations. The values of the $\tau_v/\tau_b$ ratio for actinides do not appear in Table 2, because of convergence problems in monovacancy lifetimes with the LMTO-ASA code. Finally, we have collected in Table 3 experimental positron lifetimes from ref. 17. We have fixed some conditions to select experimental values from this data base with a minimum of quality and coherence [1].

Figures 1 and 2 show the $\tau_v/\tau_b$ ratios calculated using AT-SUP (BN, LDA and GGA) and LMTO-ASA (BN and GGA) versus the atomic number, respectively. In both figures, experimental values from Table 3 have been plotted too. As it can be seen in all the curves, the $\tau_v/\tau_b$ ratio shows a periodic behaviour with the atomic number of the element. The behaviour of the $\tau_v/\tau_b$ ratio is similar to that of the positron binding energy to a monovacancy [18], and opposite to that of positron lifetime in bulk ($\tau_b$) and monovacancy ($\tau_v$), that is related to the atomic volume [1]. In the case of transition metals, where the positron localization in monovacancies is strong, the $\tau_v/\tau_b$ ratio gives the greatest values. So, metals appear at the top of the graphics in Figures 1 and 2. Indeed, $\tau_v/\tau_b$ ratio attains the maximum values when the d orbitals are semi-filled. That is, for semi-filled 3d, 4d and 5d transition metals the atomic volume ($\tau_b$ and $\tau_v$) is minimum and the positron binding energy and the ratio $\tau_v/\tau_b$ are maxima. The maximum of the positron binding energy at monovacancy when the d orbitals are semi-filled indicates that the difference between the lowest positron energy in bulk and in the defected lattice is at a maximum. Therefore, the electron-density (relative to the one of the surrounding transition elements) at the monovacancy is at a minimum, caused by the largest localization of d-electrons in semi-filled d orbitals. Therefore, under these circumstances $\tau_v/\tau_b$ ratio is maximum. In the other hand, in insulators and semiconductors the localization of the positron at monovacancies is weak. As a result, insulators and semiconductors appear at the bottom of the graphics.

Comparing Figures 1 and 2, we confirm that the ratios calculated self-consistently with LMTO-ASA code are greater than those calculated non-self-consistently with AT-SUP code. As it is known, self-consistent calculations decrease positron lifetime in bulk, and increase its value at monovacancies [5, 6]. Periodic trends appear in all the curves of Figure 1 and 2, but with different special characteristics due to the enhancement factors. As an example, in lanthanides the difference between BN and GGA is nearly constant in AT-SUP, but it increases with atomic number in LMTO-ASA. Another remarkable point is the special feature that appears around the middle of d series (4, 5 and 6 rows of the periodic table). When d orbitals get semi-filled, the $\tau_v/\tau_b$ ratio gets its maximum value (see Figures 1 and 2). In the neighbourhood of this maximum, the difference between BN and GGA ratios reduces from one row to the next one; and this happens independently of the calculation scheme used (AT-SUP or LMTO-ASA).

There are few experimental lifetimes for the elements of the periodic table. Nevertheless, the collected experimental values confirm our calculated theoretical trends, as it can be seen in Figures 1 and 2. Moreover, the ratios obtained with AT-SUP method fit better experimental values than those of LMTO-ASA. Indeed, ratios of LMTO-ASA are systematically greater than experimental ones. Finally, there are some experimental values out of the general trends (Cr, Mo, Sn and Tl). The experimental values of the $\tau_v/\tau_b$ ratio in these elements are smaller than the values of elements in their neighbourhood. This needs a further explanation:

- In the close neighbourhood of Cr ($Z = 24$, BCC structure) the following elements are located: Ti ($Z = 22$, HCP structure), V ($Z = 23$, BCC structure) and Fe ($Z = 26$, BCC structure). The experimental values follow the trend of the theoretical values. We have found only one experimental paper (see ref.

17), which measures bulk and monovacancies in Cr [19]. In the measurements performed in the cited work saturation trapping of positrons at monovacancies was not attained.

- In the surroundings of Mo ($Z = 42$, BCC structure) the elements Zr ($Z = 40$, HCP structure) and Nb ($Z = 41$, BCC structure) are placed. Their $\tau_v/\tau_b$ values fit well the theoretical ones. The Mo values obtained from experiments performed in the seventies (see ref. 17) give values of around 1,70 (1,68 [20], 1,71 [21]). However, the first value does not follow the criteria for being selected as an experimental value in ref. 17. The ones performed in the eighties (see ref. 17) are around 1,56 (1,55 [22], 1,58 [23] and 1,56 [24]). There is a value from the nineties (see ref. 17) of 1,73 [25], but vacancies are not obtained from electron irradiation experiments.

- In the close surroundings of Sn ($Z = 50$, TETRA structure) the elements Cd ($Z = 48$, HCP structure), Pb ($Z = 49$, TETRA structure) and Sb ($Z = 51$, RHOM structure) are located. The $\tau_v/\tau_b$ values of these three elements follow the trend of the theoretical values. In the case of Sn, there are two values obtained from experiments. One was performed in the seventies (see ref. 17) that gives a value of 1,39 [26], in good agreement with the theoretical value, but it does not follow the criteria for being selected as an experimental value in ref. 17. The other value, obtained in the eighties, amounts to 1,20 [27].

- Finally, close to Tl ($Z = 81$, HCP structure) are placed: Au ($Z = 79$, FCC structure), Pb ($Z = 82$, FCC structure) and Au ($Z = 83$, RHOM structure). In Au, Pb and Au, the $\tau_v/\tau_b$ values fit well the theoretical ones. In the case of Tl, there are two values obtained from experiments (see ref. 17). One was performed in the seventies that gives a value of 1,09 [28], and another from the eighties that amounts to 1,14 [29].

Summarizing, the previous data indicate that the crystals from the neighbourhood elements of Cr, Mo, Sn and Tl follow the theoretical trend, having some of them the same structure; so, one should expect these elements to follow it, too. Moreover, simulations made in metals show that relaxations are small in monovacancies, and the monovacancy lifetimes do not vary much from the non-relaxed values [30,31,32]. Therefore, the previous comments indicate that the experimental values should follow the theoretical trend, and the published experimental lifetime values of these metals are not good enough. On the other hand, the experimental data in the literature is scarce on some of those elements, and the obtained values quite different in others. Therefore, in general more experimental work is needed.

Barbiellini and coworkers found that positron lifetimes calculated with GGA approximation explain very well experimental results [5]. They also found that lifetimes obtained with BN and LDA are systematically shorter than experimental ones. However, results of BN and LDA calculated with LMTO-ASA could explain very well the experimental values using a scaling factor: 1,1 for BN and 1,21 for LDA [5]. For this reason, they pointed out that theoretical $\tau_v/\tau_b$ ratio is not very sensitive to the enhancement factor [5, 6]. Nevertheless, our calculations do not show this behaviour (see Figures 1 and 2). For proving this fact, we have made a least-squares fit between calculated and experimental positron lifetimes: $\tau_{exp} = \alpha_i \tau_i + \beta_i$ where $i$ = BN, LDA and GGA (see Table 4). Our results are similar to those of Barbiellini et al. for bulk lifetimes [5]. But, in the case of monovacancy, least-squares fit gives smaller values for $\alpha_i$ constants. So the proposed relation of proportionality is not a general law for all the calculations. In all the cases, the correlation coefficient of the fit is very high (see Table 4), so the linear relation between theoretical and experimental values is justified. However, in some cases, the values of $\beta_i$ coefficients are not negligible. For measuring the dispersion of the values, we have taken a fringe of 15 ps in both sides of the regression line, and calculated the per cent of points inside of the fringe (see Table 4). This parameter shows that the dispersion of values in monovacancy is greater than in bulk. However, this dispersion of values expresses that a general rule can not be stated for scaling positron lifetimes. So, a good relation between lifetimes of BN, LDA and GGA for one single calculation cannot be made, because, due to the effect of the enhancement factors, the changes on the electronic structure of the elements are not reflected in the $\tau_{exp}/\tau_i$ ratio as a constant scaling.

To enlighten this fact we show in Figure 3 the quotients $[(\tau_v/\tau_b)_{BN}/(\tau_v/\tau_b)_{GGA}]$ and $[(\tau_v/\tau_b)_{LDA}/(\tau_v/\tau_b)_{GGA}]$ for AT-SUP calculations as a function of the atomic number. The change of this parameter ranges between 1,12 and 0,95 for the first quotient, BN/GGA, and between 1,13 and 0,93 for the second quotient, LDA/GGA. Figure 3 shows clearly that BN, LDA and GGA approximations do not give the same value for the $\tau_v/\tau_b$ ratio. Although the deviation is not very great, they can not be neglected. Even more, the quotients plotted in Figure 3 show periodic behaviour, i.e., they are electronic structure dependent. In Figure 4, we have plotted $[(\tau_v/\tau_b)_{BN}/(\tau_v/\tau_b)_{GGA}]$ parameter calculated with AT-SUP and LMTO. In this case, the ratio values range between 1,14 and 0,91 for LMTO-ASA, and between 1,12 and 0,95 for AT-SUP. The greater differences due to the effect of the calculation method appear in lanthanides. In the case of AT-SUP calculations the positron localization at monovacancies increases in GGA with respect BN, and the opposite happens in LMTO-ASA.

So, our results show that the use of a constant scaling factor between lifetimes calculated within GGA, BN and LDA approximations is not a good assumption for improving the fit of experimental results independently of the calculation method (AT-SUP or LMTO). Therefore the $\tau_v/\tau_b$ ratio is not a good tool for making systematic comparisons.

## 4. Conclusions

We have performed an extensive self- and non-self-consistent calculations of positron annihilation lifetimes in most of the elements of the periodic table, using BN, LDA and GGA approximations. The $\tau_v/\tau_b$ ratio, as it has been shown previously for positron lifetimes and binding energies, presents a periodic behaviour. The behaviour of this ratio is similar to that of the binding energy, but it is opposite to the one of positron lifetimes.

The $\tau_v/\tau_b$ ratio has been systematically used for the study of positronic parameters, under the assumption that due to error cancellations it was a more robust quantity. Nevertheless, we have proved that the $\tau_v/\tau_b$ ratio is sensible to the enhancement factor, and that it is not possible to use a linear correction between the enhancement factors for getting positron lifetimes as has been previously proposed. This kind of ratios cannot be systematically used for making comparisons, because different features related to the enhancement factors and electronic structure appear in the calculations.

**Acknowledgements**

Technical and human support provided by IZO-SGI, SGIker (UPV/EHU, MICINN, GV/EJ, ERDF and ESF) is gratefully acknowledged. Financial support from the Spanish CICYT and Basque Government under Grants MAT2009-14398 and IT-443-10 is appreciated.

# FIGURE CAPTIONS

**Figure 1.**

Calculated and experimental $\tau_v/\tau_b$ ratios *versus* atomic number. The positron lifetimes of the elements of the periodic table have been calculated with AT-SUP using BN (blue circles), LDA (green rhombuses) and GGA (pink triangles) approximations. Experimental values are plotted using black stars.

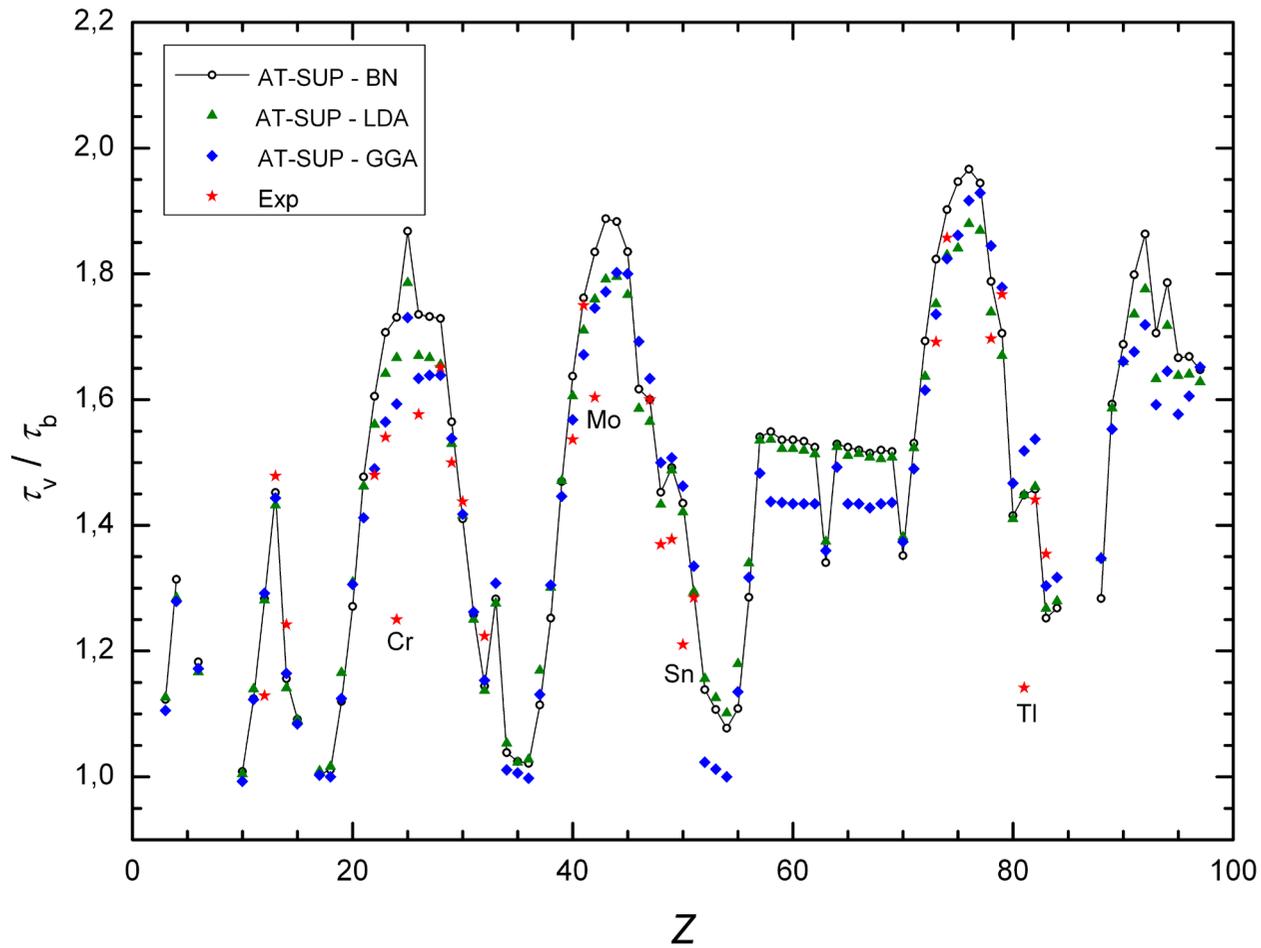

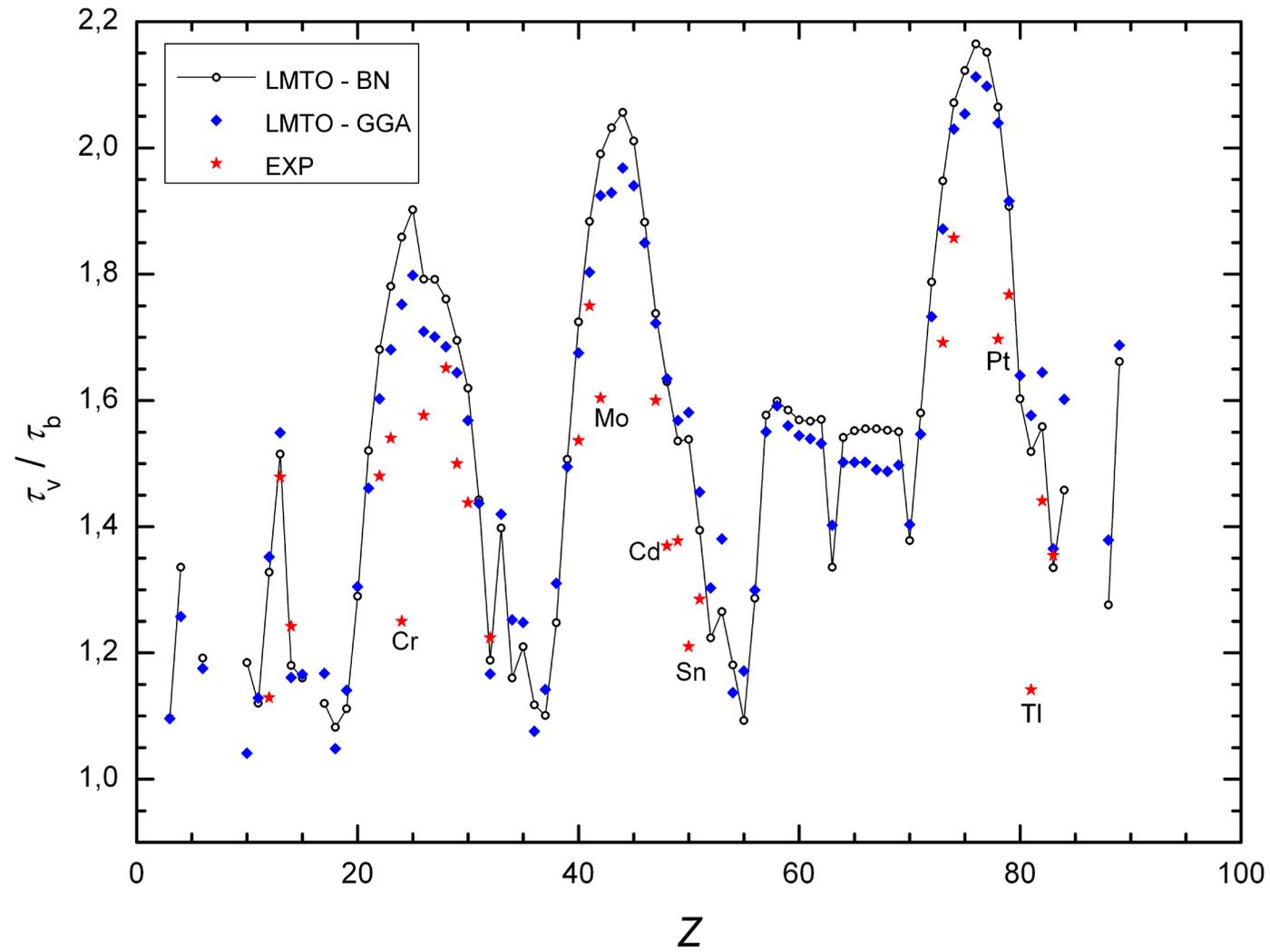

Calculated and experimental $\tau_v/\tau_b$ ratios *versus* atomic number. The positron lifetimes of the elements of the periodic table have been calculated with LMTO-ASA using BN (blue circles) and GGA (pink triangles) approximations. Experimental values are plotted using black stars.

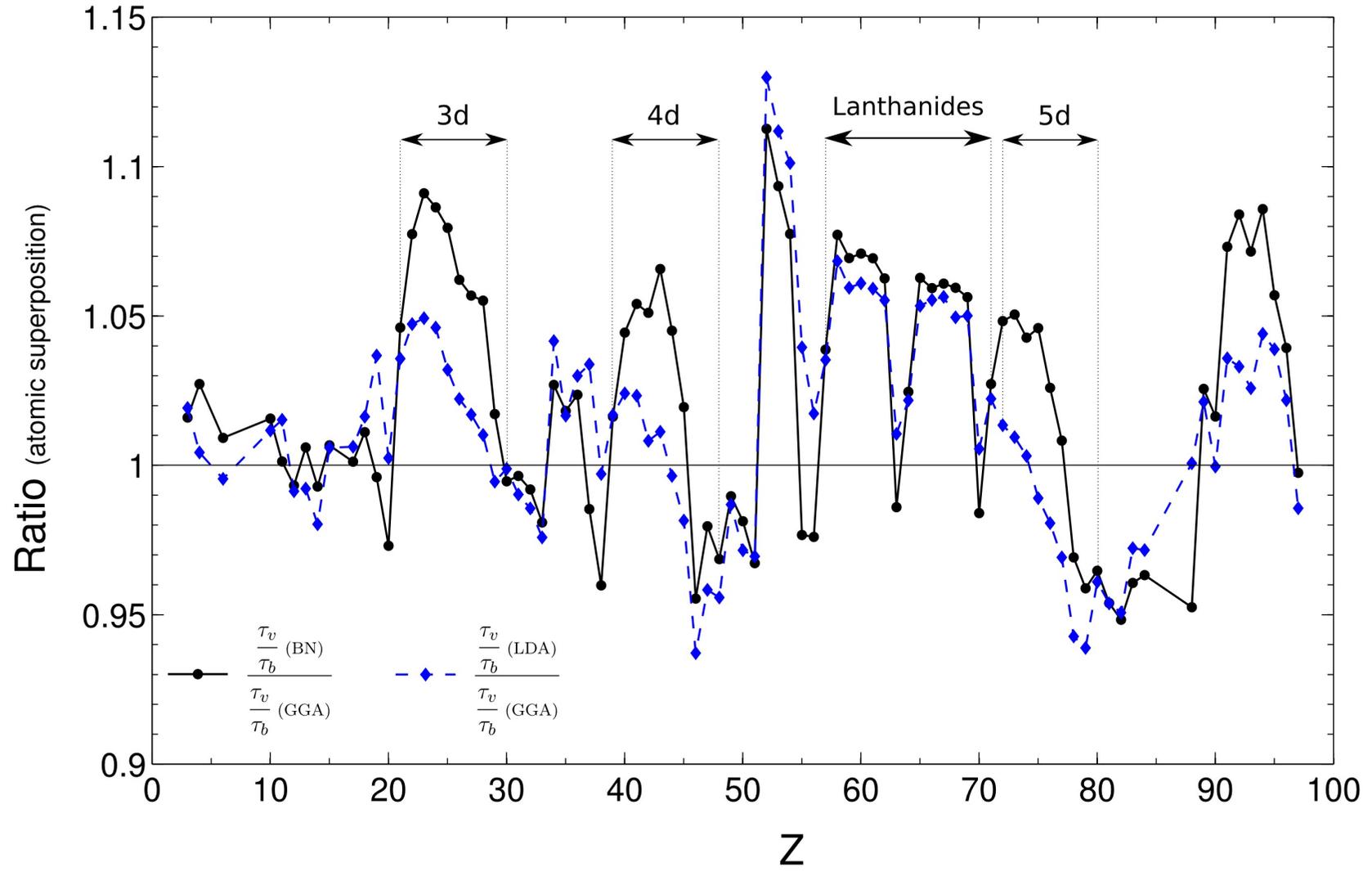

**Figure 3.**

Calculated positron lifetimes of AT-SUP expressed as $[(\tau_v/\tau_b)_{BN}/(\tau_v/\tau_b)_{GGA}]$ and $[(\tau_v/\tau_b)_{LDA}/(\tau_v/\tau_b)_{GGA}]$ *versus* atomic number.

**Figure 4.**

Calculated positron lifetimes of AT-SUP and LMTO expressed as $[(\tau_v/\tau_b)_{BN}/(\tau_v/\tau_b)_{GGA}]$ *versus* atomic number.

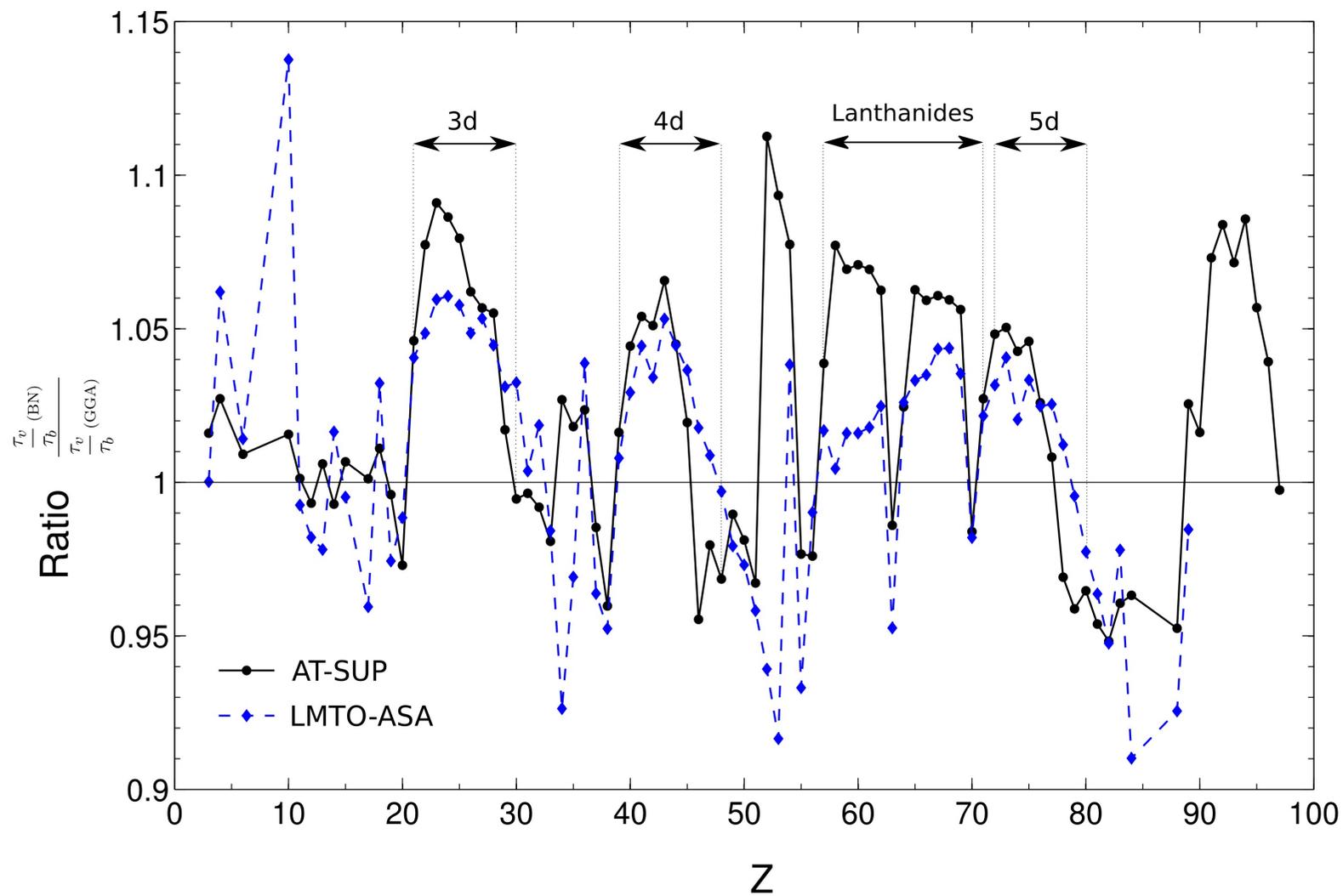

**TABLE CAPTIONS**

**Table 1.** Positron lifetime ratio, $\tau_v/\tau_b$, calculated with the AT-SUP method within BN, LDA and GGA approximations.

| H | | | | | | | | | | | | | | | | | He |
|---|---|---|---|---|---|---|---|---|---|---|---|---|---|---|---|---|---|
| — | | | | | | | | | | | | | | | | | — |
| — | | | | | | | | | | | | | | | | | — |
| **Li** | **Be** | | | | Element | | | | | | | **B** | **C** | **N** | **O** | **F** | **Ne** |
| 1,12 | 1,31 | | | | ↓ | | | | | | | — | 1,18 | — | — | — | 1,01 |
| 1,13 | 1,28 | | | | **Zr** ← $\tau_v/\tau_b$ (BN) | | | | | | | — | 1,17 | — | — | — | 1,00 |
| 1,11 | 1,28 | | | | 1,64 ← $\tau_v/\tau_b$ (LDA) | | | | | | | — | 1,17 | — | — | — | 0,99 |
| | | | | | 1,61 ← $\tau_v/\tau_b$ (GGA) | | | | | | | | | | | | |
| | | | | | 1,57 | | | | | | | | | | | | |
| **Na** | **Mg** | | | | | | | | | | | **Al** | **Si** | **P** | **S** | **Cl** | **Ar** |
| 1,12 | 1,28 | | | | | | | | | | | 1,45 | 1,16 | 1,09 | — | 1,00 | 1,01 |
| 1,14 | 1,28 | | | | | | | | | | | 1,43 | 1,14 | 1,09 | — | 1,01 | 1,01 |
| 1,12 | 1,29 | | | | | | | | | | | 1,44 | 1,16 | 1,08 | — | 1,00 | 1,00 |
| **K** | **Ca** | **Sc** | **Ti** | **V** | **Cr** | **Mn** | **α-Fe** | **α-Co** | **Ni** | **Cu** | **Zn** | **Ga** | **Ge** | **As** | **Se** | **Br** | **Kr** |
| 1,12 | 1,27 | 1,48 | 1,61 | 1,71 | 1,73 | 1,87 | 1,74 | 1,73 | 1,73 | 1,56 | 1,41 | 1,26 | 1,14 | 1,28 | 1,04 | 1,02 | 1,02 |
| 1,17 | 1,31 | 1,46 | 1,56 | 1,64 | 1,67 | 1,79 | 1,67 | 1,67 | 1,66 | 1,53 | 1,42 | 1,25 | 1,14 | 1,28 | 1,05 | 1,02 | 1,03 |
| 1,12 | 1,31 | 1,41 | 1,49 | 1,56 | 1,59 | 1,73 | 1,63 | 1,64 | 1,64 | 1,54 | 1,42 | 1,26 | 1,15 | 1,31 | 1,01 | 1,01 | 1,00 |
| **Rb** | **Sr** | **Y** | **Zr** | **Nb** | **Mo** | **Tc** | **Ru** | **Rh** | **Pd** | **Ag** | **Cd** | **In** | **β-Sn** | **Sb** | **Te** | **I** | **Xe** |
| 1,11 | 1,25 | 1,47 | 1,64 | 1,76 | 1,83 | 1,89 | 1,88 | 1,84 | 1,62 | 1,60 | 1,45 | 1,49 | 1,44 | 1,29 | 1,14 | 1,11 | 1,08 |
| 1,17 | 1,30 | 1,47 | 1,61 | 1,71 | 1,76 | 1,79 | 1,80 | 1,77 | 1,59 | 1,57 | 1,43 | 1,49 | 1,42 | 1,29 | 1,16 | 1,13 | 1,10 |
| 1,13 | 1,30 | 1,45 | 1,57 | 1,67 | 1,75 | 1,77 | 1,80 | 1,80 | 1,69 | 1,63 | 1,50 | 1,51 | 1,46 | 1,33 | 1,02 | 1,01 | 1,00 |
| **Cs** | **Ba** | **La** | **Hf** | **Ta** | **W** | **Re** | **Os** | **Ir** | **Pt** | **Au** | **β-Hg** | **Tl** | **Pb** | **Bi** | **Po** | **At** | **Rn** |
| 1,11 | 1,29 | 1,54 | 1,69 | 1,82 | 1,90 | 1,95 | 1,97 | 1,94 | 1,79 | 1,71 | 1,42 | 1,45 | 1,46 | 1,25 | 1,27 | — | — |
| 1,18 | 1,34 | 1,54 | 1,64 | 1,75 | 1,83 | 1,84 | 1,88 | 1,87 | 1,74 | 1,67 | 1,41 | 1,45 | 1,46 | 1,27 | 1,28 | — | — |
| 1,14 | 1,32 | 1,48 | 1,62 | 1,74 | 1,82 | 1,86 | 1,92 | 1,93 | 1,84 | 1,78 | 1,47 | 1,52 | 1,54 | 1,30 | 1,32 | — | — |
| **Fr** | **Ra** | **Ac** | | | | | | | | | | | | | | | |
| — | 1,28 | 1,59 | | | | | | | | | | | | | | | |
| — | 1,35 | 1,59 | | | | | | | | | | | | | | | |
| — | 1,35 | 1,55 | | | | | | | | | | | | | | | |

| **γ-Ce** | **Pr** | **Nd** | **Pm** | **Sm** | **Eu** | **Gd** | **Tb** | **Dy** | **Ho** | **Er** | **Tm** | **Yb** | **Lu** |
|---|---|---|---|---|---|---|---|---|---|---|---|---|---|
| 1,55 | 1,54 | 1,54 | 1,53 | 1,52 | 1,34 | 1,53 | 1,52 | 1,52 | 1,51 | 1,52 | 1,52 | 1,35 | 1,53 |
| 1,54 | 1,52 | 1,52 | 1,52 | 1,51 | 1,37 | 1,53 | 1,51 | 1,51 | 1,51 | 1,51 | 1,51 | 1,38 | 1,52 |
| 1,44 | 1,44 | 1,43 | 1,43 | 1,43 | 1,36 | 1,49 | 1,43 | 1,43 | 1,43 | 1,43 | 1,44 | 1,37 | 1,49 |
| **Th** | **Pa** | **U** | **Np** | **Pu** | **Am** | **Cm** | **Bk** | **Cf** | **Es** | **Fm** | **Md** | **No** | **Lr** |
| 1,69 | 1,80 | 1,86 | 1,71 | 1,79 | 1,67 | 1,67 | 1,65 | — | — | — | — | — | — |
| 1,66 | 1,74 | 1,78 | 1,63 | 1,72 | 1,64 | 1,64 | 1,63 | — | — | — | — | — | — |
| 1,66 | 1,68 | 1,72 | 1,59 | 1,65 | 1,58 | 1,61 | 1,65 | — | — | — | — | — | — |

**Table 2.**

Positron lifetime ratio, $\tau_v/\tau_b$, calculated with the LMTO method within BN and GGA approximations.

| H — — | | | | | | | | | | | | | | | | | He — — |
|---|---|---|---|---|---|---|---|---|---|---|---|---|---|---|---|---|---|
| Li 1,10 1,10 | Be 1,34 1,26 | | | | Element ↓ Zr 1,72 ← $\tau_v/\tau_b$ (BN) 1,68 ← $\tau_v/\tau_b$ (GGA) | | | | | | | B — — | C 1,19 1,18 | N — — | O — — | F — — | Ne 1,18 1,04 |
| Na 1,12 1,13 | Mg 1,33 1,35 | | | | | | | | | | | Al 1,52 1,55 | Si 1,18 1,16 | P 1,16 1,17 | S — — | Cl 1,12 1,17 | Ar 1,08 1,05 |
| K 1,11 1,14 | Ca 1,29 1,30 | Sc 1,52 1,46 | Ti 1,68 1,60 | V 1,78 1,68 | Cr 1,85 1,75 | Mn 1,90 1,80 | α-Fe 1,79 1,71 | α-Co 1,79 1,70 | Ni 1,76 1,69 | Cu 1,70 1,64 | Zn 1,62 1,57 | Ga 1,44 1,44 | Ge 1,19 1,17 | As 1,40 1,42 | Se 1,16 1,25 | Br 1,21 1,25 | Kr 1,12 1,08 |
| Rb 1,10 1,14 | Sr 1,25 1,31 | Y 1,51 1,50 | Zr 1,72 1,68 | Nb 1,88 1,80 | Mo 1,99 1,92 | Tc 2,03 1,93 | Ru 2,05 1,97 | Rh 2,01 1,94 | Pd 1,88 1,85 | Ag 1,74 1,72 | Cd 1,63 1,63 | In 1,54 1,57 | β-Sn 1,54 1,58 | Sb 1,39 1,46 | Te 1,22 1,30 | I 1,27 1,38 | Xe 1,18 1,14 |
| Cs 1,09 1,17 | Ba 1,29 1,30 | La 1,58 1,55 | Hf 1,79 1,73 | Ta 1,95 1,87 | W 2,07 2,03 | Re 2,12 2,05 | Os 2,16 2,11 | Ir 2,15 2,10 | Pt 2,06 2,04 | Au 1,91 1,92 | β-Hg 1,60 1,64 | Tl 1,52 1,58 | Pb 1,56 1,64 | Bi 1,33 1,36 | Po 1,46 1,60 | At — — | Rn — — |
| Fr — — | Ra 1,28 1,38 | Ac 1,66 1,69 | | | | | | | | | | | | | | | |

| | γ-Ce 1,60 1,59 | Pr 1,59 1,56 | Nd 1,57 1,54 | Pm 1,57 1,54 | Sm 1,57 1,53 | Eu 1,34 1,40 | Gd 1,54 1,50 | Tb 1,55 1,50 | Dy 1,56 1,50 | Ho 1,56 1,49 | Er 1,55 1,49 | Tm 1,55 1,50 | Yb 1,38 1,40 | Lu 1,58 1,55 |
|---|---|---|---|---|---|---|---|---|---|---|---|---|---|---|

**Table 3.**

Experimental positron lifetimes for bulk, $\tau_b$, and monovacancy, $\tau_v$, states, and their ratio $\tau_v/\tau_b$.

| H — — — | | | | | | | | | | | | | | | | | He — — — |
|---|---|---|---|---|---|---|---|---|---|---|---|---|---|---|---|---|---|
| Li 291 — — | Be 137 — — | | | Zn 153 220 1,44 | $\tau_b$ (ps) 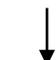 $\tau_v$ (ps) $\tau_v/\tau_b$ | | | | | | | B — — — | C 107 — — | N — — — | O — — — | F — — — | Ne — — — |
| Na 338 — — | Mg 225 254 1,13 | | | | | | | | | | | Al 165 244 1,48 | Si 219 272 1,24 | P — — — | S — — — | Cl — — — | Ar 430 — — |
| K 397 — — | Ca 230 — — | Sc 150 222 1,48 | Ti 124 191 1,54 | V 120 150 1,25 | Cr | Mn 111 175 1,58 | α-Fe 119 — — | α-Co 109 180 1,65 | Ni 120 180 1,5 | Cu 153 220 1,44 | Zn 198 — — | Ga 228 279 1,22 | Ge | As 335 — — | Se | Br | Kr |
| Rb 406 — — | Sr | Y 249 — — | Zr 164 252 1,54 | Nb 120 210 1,75 | Mo 106 170 1,60 | Tc | Ru | Rh | Pd 98 — — | Ag 130 208 1,60 | Cd 184 252 1,37 | In 196 270 1,38 | β-Sn 200 242 1,21 | Sb 214 275 1,29 | Te | I | Xe 400 — — |
| Cs 418 — — | Ba | La 241 — — | Hf | Ta 120 203 1,69 | W 105 195 1,86 | Re | Os | Ir | Pt 99 168 1,70 | Au 116 205 1,77 | β-Hg | Tl 226 258 1,14 | Pb 204 294 1,44 | Bi 240 325 1,35 | Po | At — — — | Rn — — — |
| Fr — — — | Ra | Ac | | | | | | | | | | | | | | | |

| γ-Ce | Pr | Nd | Pm | Sm 199 — — | Eu | Gd 230 — — | Tb | Dy | Ho | Er | Tm | Yb | Lu |
|---|---|---|---|---|---|---|---|---|---|---|---|---|---|

**Table 4.**

Least-squares fit between theoretical and experimental positron lifetimes ratios, correlation factor, and per cent number of points inside a fringe of 15 ps in both sides of the regression line.

|  |  | AT-SUP |  |  | LMTO |  |  |
|---|---|---|---|---|---|---|---|
|  |  |  | $r$ | % |  | $r$ | % |
| Bulk (44 data points) | BN | $1{,}15\,\tau_{BN} - 10$ | 0,96 | 86 | $1{,}13\,\tau_{BN} - 4$ | 0,96 | 82 |
|  | LDA | $1{,}30\,\tau_{LDA} - 13$ | 0,96 | 84 |  |  |  |
|  | GGA | $0{,}93\,\tau_{GGA} + 10$ | 0,98 | 84 | $0{,}98\,\tau_{GGA} + 11$ | 0,99 | 89 |
| Monovacancy (26 data points) | BN | $0{,}97\,\tau_{BN} + 7$ | 0,90 | 65 | $0{,}98\,\tau_{BN} - 5$ | 0,92 | 73 |
|  | LDA | $1{,}14\,\tau_{LDA} - 2$ | 0,90 | 73 |  |  |  |
|  | GGA | $0{,}83\,\tau_{GGA} + 24$ | 0,88 | 65 | $0{,}90\,\tau_{GGA} + 7$ | 0,91 | 69 |